\documentclass[showpacs,nofootinbib,superscriptaddress]{revtex4}
\usepackage{graphicx}
\usepackage{color}
\def\be{\nopagebreak[3]\begin{equation}}
\def\ee{\end{equation}}
\def\ba{\nopagebreak[3]\begin{eqnarray}}
\def\ea{\end{eqnarray}}

\newcommand{\teta}{\rlap{\lower2ex\hbox{$\,\tilde{}$}}\eta{}}

\begin{document}
\preprint{\vbox{\baselineskip=12pt \rightline{IGPG-07/6-?}
\rightline{gr-qc/yymmnnn} }}
\title{Black hole radiation spectrum in LQG: Isolated Horizon framework}

\author{Jacobo D\'\i az-Polo}
\email{Jacobo.Diaz@uv.es}\affiliation{Departamento de Astronom\'\i a
y Astrof\'\i sica, Universidad de Valencia, Burjassot-46100,
Valencia, Spain}

\author{Enrique Fern\'andez-Borja}\email{Enrique.Fernandez@uv.es}
\affiliation{Departamento de F\'\i sica Te\'orica and IFIC, Centro
Mixto Universidad de Valencia-CSIC. Universidad de Valencia,
Burjassot-46100, Valencia, Spain}

\begin{abstract}

Recent detailed analysis within the Loop Quantum Gravity calculation of
black hole entropy shows a stair-like structure in the behavior of
entropy as a function of horizon area. The non-trivial distribution of the degeneracy
of the black hole horizon area eigenstates is at the origin of this behavior.
This degeneracy distribution is analyzed and a phenomenological model is put forward to study the implications of this distribution in the black hole radiation spectrum. Some qualitative quantum effects are obtained within the Isolated Horizon framework. This result provides us with a possible observational test of this model for quantum black holes.

\end{abstract}

 \maketitle

\section{Introduction}
Black hole radiation is one of the most outstanding features of quantum gravity. Its study is one of the pivotal points of any approach to such a theory. The semiclassical behavior of this effect was derived out by Hawking \cite{Hawking}, but some kind of fine structure is expected to arise within a full theory of quantum gravity.

In Loop Quantum Gravity (LQG) the best established framework for the study of black holes is the {\emph{Isolated Horizon}} quantization proposal \cite{ABCK}. This approach provides a description of the detailed quantum configurations of a black hole in terms of quantum geometrical states. Considering these states, one can obtain a microscopical description of the black hole entropy \cite{ABCK,Dom:Lew,Meiss,GM,CDF}.

Recently, an explicit computational counting of those states
has been carried out \cite{CDF,CDF1}, allowing to find a stair-like behavior in entropy as a function of area, that had remained hidden until then. The non-trivial distribution of the black hole horizon area eigenstate degeneracy is found to be at the origin of this behavior. More specifically, the most degenerate eigenstates accumulate around certain evenly spaced values of area. The implications of this structure can be extended to other black hole properties. In particular, in this paper we will focus on the qualitative picture for the black hole radiation spectrum that one can extract from this feature.

Within LQG, the dynamics producing the black hole radiation process is not well established (some recent works about this topic are \cite{ashtekar}); this allows one to study the spectroscopy only at the kinematical level. Some works have been done in this direction \cite{Krasnov, Rovelli} and also \cite{Ansari}, but on the basis of certain additional assumptions. In this last reference, it was claimed that the Isolated Horizon framework does not provide any quantum gravity imprint in the black hole radiation spectrum. Here we shall explain how this framework provides, in fact in a natural way, a quantum imprint in the black hole radiation spectrum at the kinematical level, without any modification of the initial hypothesis. Furthermore, we will obtain a picture compatible, in some sense, with Bekenstein's conjecture \cite{beken2} as a first approximation. Substantially different qualitative features are expected to arise, however, when a more complete analysis of the full set of states in LQG is considered. These results could open a window for some hypothetical observational tests of the model, although one should be cautious about this possibility, as dynamics could introduce important modifications.

\section{Isolated Horizon Canonical Quantization and Explicit state counting}
Let us briefly review the Isolated Horizon quantization framework. In this approach, black holes are treated in an {\it effective} way, since the Isolated Horizon is introduced from the outset as an inner boundary of the spacetime manifold before the canonical quantization procedure is carried out.
Given a fixed area $A_0$ for the black hole horizon, the states that
have to be considered arise from a punctured spherical surface,
where punctures carry some `quantum' labels that must satisfy two constraints: compatibility with the horizon area and a quantum analogue of Gauss-Bonnet theorem that ensures the spherical topology of the horizon. For a  detailed description of such procedure, see \cite{ABCK}.

LQG has a quantization ambiguity due to the presence of a free real
parameter $\gamma$, the Barbero-Immirzi (BI) parameter \cite{BI}, which gives rise to inequivalent quantum theories.
The entropy calculation in LQG recovers
the functional relation of entropy vs. area up to this parameter.  The
standard procedure is to impose the Bekenstein-Hawking entropy-area law \cite{Hawking, BH} for large black holes in order to fix the value of $\gamma$. In addition, a logarithmic correction with a $-1/2$ coefficient is obtained; this term turns out to be independent of the value of the BI parameter.

There is another possible source of ambiguity, since there are two possible choices of labels to describe the black hole horizon states which contribute in the entropy
calculation (for details see \cite{Dom:Lew,Meiss,GM}).
The main difference between both models is that they give rise to different values for the BI-parameter ($\gamma$) as a result. Which is the proper choice of labels that should be considered for the computation of black hole entropy is, by itself, a very interesting (still open) problem, that involves subtleties in the splitting of the horizon and bulk degrees of freedom. Nevertheless, the degeneracy distribution that we are going to analyze here was found in \cite{CDF1} to be independent on this choice of labels, so we are not concerned with this problem now. Furthermore, this model independence gives some robustness to the result and seems to point out that this effect is an intrinsic feature of the Isolated Horizon framework.

Let us then summarize the results that were obtained in \cite{CDF,CDF1} with the exact counting of states.
In order to make a microcanonical analysis, representative states of the black hole horizon
are assumed to have two labels $(j,m)$ assigned to each puncture,
where $j$ is a
typical spin-like number $j=\{\frac{1}{2}, 1, \frac{3}{2}, ...\}$ that carries information about area,
and $m$ is its corresponding projection, satisfying $m=\{-j, -j+1, ..., j\}$, which carries information about the curvature of the horizon (through the Isolated Horizon boundary conditions).
An explicit choice between the two possible assignments of labels for the horizon states has been made at this point,
attending to consistency with the microcanonical framework (the other possible choice only takes into account
the $m$ labels, but considering also $j$ allows us to have a definite value of area for each microstate, which
is needed to treat the system as a proper microcanonical ensemble). Anyway, as pointed out above, this choice does not affect the results we are going to analyze. Thus, the two constraints
that states must satisfy take the explicit form:
\be
\label{ligarea}
A_0-\delta\leq 8\pi \gamma \ell_P^2\sum_{p=1}^N{\sqrt{j_p(j_p+1)}}\leq A_0+\delta,
\ee
where $\delta$ is an area of the order of Planck scale and $p$ labels the different punctures, i.e., the sum of the area contributions
of each puncture equals the total horizon area (within a certain tolerance interval), and
\be\label{ligproy}
\sum_{p=1}^N{m_p=0},
\ee
which implements the condition of a spherical horizon.
With this definition of states, the value $\gamma=0.274$ for the BI-parameter
is obtained \cite{GM,CDF,Krip}.

When a {\it{brute force}} analysis of states is performed,
what it is being counted, roughly speaking, is all the different ways to combine the labels
$(j,m)$ (for any possible total number of punctures) that are consistent with the above constraints and with the
distinguishability of punctures as it is explained in
\cite{CDF}. The logarithm of this number is then taken in order to obtain the corresponding
entropy. When this analysis is done for increasing values of the horizon area
$A_0$, the asymptotic linear behavior of entropy is rapidly recovered, even for small values
of $A_0$, far below the large area limit in which the Isolated Horizon framework of LQG was originally formulated. Moreover,
we can verify the emergence of a logarithmic correction, due to the introduction of the constraint (\ref{ligproy}), with a
coefficient of value $-1/2$. These facts give one some confidence
in the relevance of performing such a counting even though, due to running time computational limitations, one is restricted to work in a small horizon area regime (as the number of states to be counted grows exponentially with area, it is not possible to compute arbitrarily large areas in a reasonable time).
Of course, the results obtained
in this regime cannot be lightly extrapolated to the large area regime, and the algorithm should be optimized in order to reach larger values of area. Nevertheless, to explore the predictions of the LQG framework in this Planck scale limit is, by itself, an interesting subject which could help us to gain some insight into the behavior of microscopical black holes.

\begin{figure}
\begin{center}
\includegraphics[width=8cm]{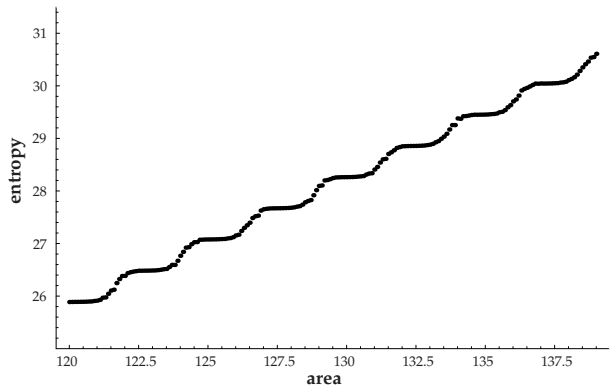}
\end{center}
\caption{Plot of the results for the entropy as a function of area (in Planck units) obtained with the computational counting.
The stair-like behavior is observed.}
\label{escalera}
\end{figure}

There is a surprising additional feature that has been obtained.
A staircase structure (Fig. \ref{escalera}) is found when one
plots the entropy as a function of the horizon area \cite{CDF1}.
Entropy increases with the area by discrete equidistant steps.
This means that, in some sense, entropy is discretized within the LQG approach. The area interval where the entropy is almost constant (the width of each step in the stair) is given by $\Delta A \approx 2.41...$.

At this point one may ask whether this behavior would be recovered with the other possible choice of states and counting prescription, as given in \cite{Dom:Lew}. The answer is affirmative, as was pointed out in \cite{CDF1}, so
the same entropy staircase is obtained within the model of Domagala and Lewandowski,
but this time with a step width of $\Delta A_{\rm DL} \approx 2.09...$.

The relevant fact is that the value of $\Delta A$ turns out to be
proportional to the value of $\gamma$ obtained in each model:
\be\label{Prop}
{\Delta A = \chi\gamma\ell_P^2},
\ee
where $\chi\approx 8\ln{3}$ (although still there is not any analytic proof for $\chi$ to take this value, it has been recently found \cite{Hanno, ADF} that the numerical coincidence holds, at least, up to an accuracy of a part in $10^4$).

The fact that the stair-like behavior is independent of the choice of states is
somewhat unexpected and seems to point out that this
result is an artifact of neither the definition of horizon states nor
the computational counting of them \cite{Dom:Lew,GM}.
In fact, the staircase has its origin in the LQG area spectrum itself \cite{QTGI}
and the structure of the eigenstate degeneracy distribution obtained when imposing the constraint (\ref{ligproy}).

\begin{figure}
\begin{center}
\includegraphics[width=8cm]{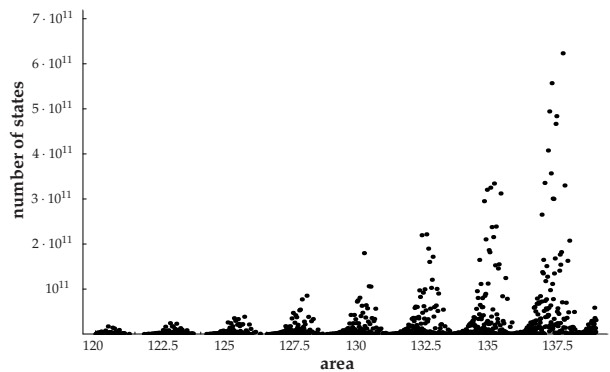}
\end{center}
\caption{Degeneracy (number of different horizon states in each area interval of $0.01\ell_P^2$) is plotted.
States accumulate around some equidistant area values.}
\label{histograma}
\end{figure}

As one can notice from (Fig. \ref{histograma}), the degeneracy of horizon area eigenstates is distributed following
a structure with bands, where on each `band' there is a particular peak on the
number of states, around which most of the states accumulate. The separation between
peaks is approximately equidistant in area. The area gap between two consecutive
peaks of degeneracy is just $\Delta A$.  The contribution to the entropy from the states at the peaks
is several orders of magnitude higher than the contribution of the rest of the spectrum (states out of the peaks).
This behavior of the black hole state degeneracy provides an explanation to the entropy staircase\footnote{Considering all states contained in a certain area interval, in order to compute microcanonical entropy, gives rise to the discrete jumps when, while displacing it, this interval changes from containing one peak to the next one (provided that the width $2\delta$ of the interval is chosen to be equal to the spacing $\Delta A$ between peaks). For the area regions at which the interval keeps containing the same peak the entropy remains constant.}.
Here, however, we want to extend our analysis to the implications of this distribution of states for black hole radiation. We will do this in the next section.

\section{Qualitative spectroscopical analysis}

The issue of obtaining a qualitative picture for the black hole radiation spectrum arising when quantum effects are considered was addressed by Bekenstein on the basis of some heuristic arguments. Let us briefly review the conjectured results (all the
details can be found in \cite{beken2}). The adiabatic invariance properties of the horizon area
lead to its quantization in the sense of Ehrenfest's principle
\be\label{generalbek}
a_n=f(n) \hspace{1cm} n=1,2,3,\ldots ,
\ee
where $a_n$ is the horizon area spectrum and $f(n)$ is
a positive, monotonically increasing function.
On the basis of the analysis of a quantum version of a
Christodoulou process \cite{cristo}, an evenly spaced spectrum of the type
\be
\label{bekspect} a_n=4n\ell_p^2\ln{k}
\ee
is proposed, where $k$ is a positive integer and the $4\ln{k}$ factor arises from consistency with the entropy-area law.
This spectrum is based on `semiclassical' considerations and
Bekenstein's conjecture establishes that it holds for all area regimes.

Taking into account the minimal change in area, and considering the corresponding
transition between two consecutive area eigenstates, one can give the
frequency of the transition assuming the Bohr's correspondence
principle
\be \label{bekfrec} \omega_0=\frac{\ln{k}}{8\pi M}.\ee
This was done by Bekenstein and Mukhanov in \cite{bek-muk}. They considered this frequency as the
minimal emission frequency; black hole radiation should occur in integer multiples of this
fundamental frequency, giving rise to a discrete spectrum.

Here we are going to do some similar qualitative considerations but on the basis of the results obtained from the Isolated Horizon quantization framework in Loop Quantum Gravity. This framework is valid for (locally) stationary horizons, and the dynamics that would be responsible of the radiation process is not implemented. However, we can make a spectroscopical analysis at the kinematical level, considering transitions between stationary initial and final horizon states (in the spirit of atomic spectroscopy with the time independent Schrodinger equation).

In a first step we are going to neglect the detailed structure of the spectrum and consider only the peaks as allowed values of area. This will give a picture compatible with the Bekenstein and Mukhanov works in a first approximation. However, some qualitative differences will arise when considering the black hole state spectrum a little more in detail.
Of course, a more comprehensive study considering
the structure of the bands in full detail is needed in order to unravel the
precise behavior of black hole radiation within this framework, but we will leave this
complete study for further investigation.

Therefore, in a first approximation we can consider the relation
\be
\label{CDF}{\Delta A = \gamma\,\chi\,\ell_P^2}
\ee
as defining an equidistant area spectrum for the horizon. This is analogous to considering that the black hole has permissible states only for those values of area corresponding to the peaks of the
bands.
This equidistant area spectrum will provide us with a fundamental frequency, corresponding to transitions between two consecutive peaks, and allowed transitions (decays) would imply integer multiples of this frequency, in the spirit of Bekenstein and Mukhanov's work. We can analogously extract a frequency from our fundamental area interval (\ref{CDF}). We then obtain
\be\label{lqgfrec}
\omega_0^{\rm LQG}=\frac{\gamma\chi}{32\pi M}.
\ee
Then, within this approximation, one recovers in a very precise way the Bekenstein-Mukhanov's picture for black hole radiation arising from LQG in a natural way, but with a slightly different value for $\omega_0$ as
they do. We have an equidistant area spectrum with the only difference of the area gap, so an analogous analysis can be made.

However, we do not have states only at the peaks. The values of area corresponding with the peaks are statistically preferred, but not the only allowed values. Thus, there is a very important difference between the physical conclusions that one can extract from the Bekenstein-Mukhanov picture and from the simplified model considered here. While the former implies the discreteness of the black hole radiation spectrum, the latter is to be seen only as a simplification of a much more intricate structure. We can only say that there are some statistically preferred transitions, in the sense of Fermi's Golden Rule. Then, our frequency, and its integer multiples, are not meant to be the only emitted frequencies, but some brighter lines sticking out from an almost continuous spectrum. This brighter lines would not be thin and sharp, but rather broad and smooth, as one can expect from the structure of (\ref{histograma}). The exact details of this broadening should be obtained when performing the complete spectroscopical analysis, which will be carried out elsewhere.

Moreover, the fact that the frequencies of this brighter lines are proportional to $\gamma$ could open a window to the observational determination of the value of this parameter, and furthermore, since the relation (\ref{Prop}) is `model independent', also of the appropriate choice of horizon states. Besides, given that the fundamental frequency
$\omega_0$ is near the peak of the black body spectrum \cite{bek-muk}, this
effect should, in principle, be observable. Though in principle, one should be still quite far from the possibility of performing such observations, it is worth to comment on such a hypothetical contact point with observation.

\section{Remarks and conclusion}

Let us conclude with some remarks. In our analysis we are assuming that a black hole can undergo a transition from any configuration to any other one with the only condition that the final state belongs to a lower area band (transitions within the same band would imply no change in entropy, and then would be reversible processes that can not be seen as producing radiation). This includes changes in the total number of punctures (both increasing and decreasing) as well as changes in the values of the labels of punctures (again, both increasing and decreasing).

It is important to distinguish our model from others that have appeared in the literature, such as \cite{Ansari}, where brighter lines on a continuous spectrum are also obtained, but on the basis of quite different considerations. In that work it is argued that the Isolated Horizon framework does not produce any quantum gravity imprints in the black hole radiation spectrum, in contrast with the present result. Furthermore, in \cite{Ansari} decays of the black hole are supposed to be produced by sequences of individual transitions of punctures, in which each of the involved punctures looses part of its contribution to the area by decreasing the value of its quantum labels. This is an additional assumption introduced by hand, as the dynamics of the transition process is unknown.
Here we are taking into account all possible transitions in a much more general way, as explained above.

In this sense, we are considering the black hole as a system which can take some microscopical configurations (microstates in the statistical sense) that are equally probable. Punctures have not an individual identity and are not to be seen as a kind of particles \cite{LQG FAQ}. A (quantum) black hole is not a system that can be decomposed into these individualities, it only makes sense as a whole. Transitions of the black hole should then be considered in this general way, between any two possible microstates, and should not be restricted only to those implying individual decays of punctures.

In conclusion, we have argued that the Isolated Horizon framework in LQG gives rise to some quantum effects in the black hole radiation spectrum at the kinematical level for a low horizon area regime. These effects can accommodate, in a first approximation, the equidistant area spectrum for black holes conjectured by Bekenstein with the corresponding entropy discretization. Nevertheless, the physical consequence that one can extract when studying the spectroscopy is not the discretization of the radiation spectrum and the appearance of a minimum emission frequency (as in the Bekenstein-Mukhanov scenario).
The imprint of quantum gravity effects in Hawking radiation (for microscopic black holes) within this framework is manifested in the emergence of some equidistant brighter lines over a continuous background spectrum. One can only speculate that such an effect could be searched in some microscopic black hole candidates (such as GRB's) \cite{GRB}. The hypothetical appearance in their spectrum of these qualitative effects could be an indication of the validity of the model presented here and the frequency of these brighter lines (specifically, the fundamental frequency $\omega_0$, if found) could also be used to give an (indirect) observational determination of the Barbero-Immirzi parameter.

We specially thank A.~Corichi for his advise and encouragement.
We also thank J.~Olivert, J.~D.~Bekenstein, G.~Mena-Marug\'an and H.~Sahlmann for discussions.
We finally thank A.~Corichi, J.~A.~de~Azc\'arraga, I.~Agull\'o and I.~Garay for useful
comments and suggestions and a careful reading of the manuscript.
This work was in part supported by ESP2005-07714-C03-01 and FIS2005-02761 (MEC) grants.
J.D. thanks MEC for support through the FPU (University Personnel Training) Program.

\end{document}